\documentclass[useAMS,usenatbib]{mn2e}
\usepackage{graphicx,epsfig,longtable}
\usepackage{color}
\usepackage{amsmath}
\usepackage{tabularx} 
\usepackage{hyperref}

\newcommand{\beqy}{\begin{eqnarray}}
\newcommand{\eeqy}{\end{eqnarray}}
\newcommand{\bmlet}{\begin{subequations}}
\newcommand{\emlet}{\end{subequations}}

\def\gsimeq{\,\,\raise0.14em\hbox{$>$}\kern-0.76em\lower0.28em\hbox
{$\sim$}\,\,}
\def\lsimeq{\,\,\raise0.14em\hbox{$<$}\kern-0.76em\lower0.28em\hbox
{$\sim$}\,\,}
\newcommand{\Msun}{$M_{\odot}$}
\newcommand{\aap}{A\&A}
\newcommand{\apjl}{ApJL}
\newcommand{\apjs}{ApJS}
\newcommand{\apj}{ApJ}

\newcommand{\prc}{Phys. Rev. C}
\newcommand{\prl}{Phys. Rev. Lett.}
\newcommand{\mnras}{MNRAS}

\newcommand{\physrep}{Phys. Rep.}
\newcommand{\prd}{Phys. Rev. D}
\newcommand{\chem}[2]{$\rm{}^{#1}\kern-0.8pt#2$}

\title[Actinide production in neutrino-driven winds]
 {Solar r-process-constrained actinide production in neutrino-driven winds of supernovae}

\author[S. Goriely et al.]{S.~Goriely$^1$ and H.-Th.~Janka$^2$ \\
  $^1$Institut d'Astronomie et d'Astrophysique, Universit\'e Libre de Bruxelles,  CP 226, 1050 Brussels, Belgium \\
  $^2$Max-Planck-Institut f\"ur Astrophysik, Postfach 1317, 85741 Garching, Germany \\
}

\date{Released 2016 Xxxxx XX}

\pagerange{\pageref{firstpage}--\pageref{lastpage}} \pubyear{2016}

\begin{document}
\label{firstpage}
\maketitle

\begin{abstract}
Long-lived radioactive nuclei play an important role as nucleo-cosmochronometers
and as cosmic tracers of nucleosynthetic source activity. In particular nuclei
in the actinide region like thorium, uranium, and plutonium can testify to the 
enrichment of an environment by the still enigmatic astrophysical sources
that are responsible for the production of neutron-rich nuclei by the rapid 
neutron-capture process (r-process). Supernovae and merging neutron-star (NS)
or NS-black hole binaries are considered as most likely sources of the r-nuclei.
But arguments in favour of one or the other or both are indirect and make use of
assumptions; they are based on theoretical models with remaining simplifications
and shortcomings. An unambiguous observational determination of a production
event is still missing. In order to facilitate searches
in this direction, e.g.\ by looking for radioactive tracers in stellar 
envelopes, the interstellar medium or terrestrial reservoirs, we provide 
improved theoretical estimates and corresponding uncertainty ranges for
the actinide production ($^{232}$Th, $^{235,236,238}$U, $^{237}$Np, 
$^{244}$Pu, and $^{247}$Cm)
in neutrino-driven winds of core-collapse supernovae. Since state-of-the-art
supernova models do not yield r-process viable conditions ---but still lack,
for example, the effects of strong magnetic fields--- we base our investigation
on a simple analytical, Newtonian, adiabatic and steady-state wind model and
consider the superposition of a large number of contributing components, whose
nucleosynthesis-relevant parameters (mass weight, entropy, expansion time scale, 
and neutron excess) are constrained by the assumption that the integrated wind 
nucleosynthesis closely reproduces the solar system distribution of r-process 
elements. We also test the influence of uncertain nuclear physics.
\end{abstract}


\section{Introduction}
\label{sect_intro}

Actinides play an important role in astrophysics in different ways. 
First, the use of \chem{232}{Th},
\chem{235}{U} and \chem{238}{U} to estimate astrophysical ages has a long history, a
milestone of which is the much celebrated piece of work of \citet{fowler60}. 
For long, the field of nucleo-cosmochronology that emerged
from this paper has been aiming at the determination of the age $T_{\rm nuc}$ of
the nuclides from abundances in the material making up the bulk of the solar
system.
The astrophysical importance of Th and U was enhanced further with the 
observation of Th and U in some very metal-poor stars \citep{sneden96,cayrel01}. 
These measurements raised the hope of 
a possible nuclear-based evaluation of the age of individual stars other than the Sun.

Additional recent observational advances have triggered substantial interest in other
actinides that are shorter-lived than Th and U. This comes about following the
measurement of the Galactic Cosmic Ray (GCR) abundances of the $Z > 70$ elements,
including the actinides, with unprecedented resolution, using the Trek detector \citep{westphal98}. 
Further significant progress is expected by the determination of the GCR
abundances of the actinides Th, U, Pu and Cm both relative to each
other and relative to the Pt-group of elements. Precise abundance measurements of this type
would  yield an estimate of the time elapsed between the nucleosynthesis
of the GCR actinides and their acceleration to GCR energies (the
GCR actinide propagation time after acceleration is very short, i.e of the order of 2 My).
Hence, they would help determining whether GCRs were accelerated out of fresh
ejecta of the astrophysical r-element sources (supernovae or neutron-star mergers), 
superbubble material, or old, well-mixed galactic material. 

Also attempts to measure the $^{244}$Pu content of the local interstellar medium
should be mentioned here. Such measurements may allow for conclusions on the
frequency of the astrophysical events producing actinides. At
present, this can be done through the analysis of dust grains of identified interstellar
origin recovered in deep-sea reservoirs \citep[sediments and FeMn crusts;
e.g.][]{paul01,wallner15} and by the recovery of $^{60}$Fe incorporated in
fossil biogenic samples \citep{bishop11,ludwig16,witze13}. Marine sediments were also analysed for a $^{244}$Pu
signal associated with measurements of $^{60}$Fe in a ferromanganese 
crust \citep{wallner04,raisbeck07}, which 
has been interpreted as product of a supernova close to the solar system about 
2.2\,My ago \citep{knie99,knie04,fitoussi08}.

For all these reasons, there is an obvious need to provide detailed theoretical
estimates of the 
possible stellar production of actinides with half-lives in excess of typically $10^6$~y. 
Most importantly, the uncertainties in these predicted abundances should be evaluated as well.
While the actinides are clearly produced by the rapid neutron-capture process (or r-process) 
of stellar nucleosynthesis, the site(s) of this nucleosynthesis process have not been
unambiguously identified yet \citep{arnould07}.

Two possible astrophysical sites of r-process production are discussed:
supernovae and neutron-star mergers. In the case of supernova explosions a variety
of scenarios for r-element creation by primary and secondary processes were
proposed. Neutron-rich jets in rare, magnetohydrodynamic explosions of rapidly 
rotating stars \citep[e.g.][]{winteler12} and neutron production by neutrino 
reactions in the helium layer of compact, metal-poor exploding stars 
\citep{banerjee11} were considered as being potentially responsible for an
early enrichment of the young Galaxy with r-process matter. However, supernovae
could be a major or even dominant source of r-process elements only if the 
so-called neutrino-driven wind, a low-mass baryonic outflow from newly formed
neutron stars \citep[e.g.][]{woosley94,qian96,hoffman97}, were able to provide 
an r-process viable environment \citep{argast04}. 
Self-consistent models, however, do not only
yield wind entropies that are too low to enable a strong r-process that could
make lanthanides and actinides \citep[e.g.][]{taka94,witti94,roberts10}, but
state-of-the-art supernova models with high-fidelity neutrino transport 
also yield proton-rich conditions instead of neutron excess in the wind ejecta
\citep{hudepohl10,fisch10,janka12,mirizzi16}.

Recently, growing attention has been paid to mergers of binary neutron stars (NS-NS) 
and neutron star-black hole (NS-BH) systems
as possible r-process sites \citep{lattimer76,lattimer77,eichler89}, following 
the confirmation by hydrodynamic simulations that non-negligible amounts of matter, 
typically about $10^{-3}$ to several $10^{-2}$\Msun, can be ejected 
\citep[e.g.][]{rosswog99,frei99,arnould07,metzger10,roberts11,goriely11,korobkin12,bauswein13,goriely13,wanajo14,perego14,just15,seki15,radice16}.
In contrast to the supernova case, investigations with growing
sophistication have so far supported NS merger ejecta as viable sites 
for strong r-processing \citep[e.g.][for steps towards including neutrino 
effects]{wanajo14,seki15,goriely15a,roberts16}.
In particular, comprehensive nucleosynthesis calculations \citep{just15,martin15} 
show that the combined contributions from both the dynamical (prompt) ejecta 
expelled during NS-NS or NS-BH mergers, and the 
neutrino and viscously driven outflows generated during the post-merger 
remnant evolution of relic NSs or BH-torus systems can lead to the 
production of 
r-process elements from  $A \ga90$ up to thorium and uranium with an 
abundance distribution that reproduces extremely well the solar distribution, 
as well as the elemental distribution observed in low-metallicity stars 
\citep{roederer11,roederer12}. 

Despite a roughly 1000 times lower NS merger rate compared to the core-collapse
supernova rate, NS-NS/BH mergers could still explain the total amount of r-material
in the Galaxy because the amount of ejected r-process-rich matter per NS merger is
roughly 1000 times higher than the potential contribution from the neutrino-driven
wind of a supernova. Based on observed binary pulsars and
population synthesis calculations, the NS-NS merger rate is currently 
estimated to be within the plausible range of 3 to 190~Myr$^{-1}$ 
\citep[e.g.][]{kim10,dom12,vangioni15}.
Initially there were concerns that the low rate of compact star mergers
(with their correspondingly larger ejecta masses per event) in addition to the
long time delay of binary mergers after the preceding, iron-producing supernovae
would be incompatible with the Galactic enrichment history as deduced from
observations of (ultra-)metal-poor stars \citep{argast04}. However, more recent
models of the chemical evolution of the Milky Way show a much more 
promising situation. Taking into 
account a short-lived binary component (with inspiral times of less than
100~Myr), incomplete Galactic mixing and/or the sub-halo merger history
of the Milky Way, these new studies conclude that double compact star mergers 
might indeed be the major producers of r-process elements and might even be
responsible for the enrichment of metal-poor stars
\citep{mat14,kom14,mennekens14,shen14,voort14,vangioni15,wehmeyer15,ishimaru15}.
Although these investigations, which are based on largely different modeling
approaches, ranging from traditional, highly simplified 
box models to hydrodynamical simulations,
do not agree in all quantitative details of the picture, the majority of them
suggests that compact binary mergers as r-process sites are better compatible
with the considerable event-to-event scatter of the r-process abundances in
metal-poor stars.
 
If NS-NS/BH mergers are indeed the main cosmic sources of heavy r-process matter,
the rarity of such events would leave little hope for discovering on Earth
larger amounts of radioisotopes of cosmic origin like $^{244}$Pu. 
In fact, a recent
measurement found a very low $^{244}$Pu abundance in deep-ocean reservoirs,
about two orders of magnitude lower than expected from continuous production by
frequent sources like supernovae~\citep{wallner15}. 
In combination with abundance measurements for the
Early Solar System this experimental result was therefore interpreted as strong
evidence for the origin of $^{244}$Pu from compact binary mergers \citep{hotokezaka15}.

However, more experimental and observational confirmation with better 
statistics and based on
alternative reservoirs is highly desirable to consolidate the picture 
suggested by the existing measurements, which are subject to considerable
uncertainties associated with limited statistics and our incomplete knowledge
of the probability with which supernova-made nuclei make their
way to Earth and finally end up in the investigated sample material. Such caveats
certainly justify ongoing efforts to directly identify supernovae as cosmic sources
of r-process matter or, alternatively, to derive increasingly stronger bounds
to the supernova origin of this matter by searching for r-nuclei in the envelopes
of supernova companion stars and for radioisotopes in dated terrestrial sediments.
It should be kept in mind that without any unambiguous identification of
an r-process source by the in-situ detection of r-nuclei our understanding of
the astrophysical origin of this nucleosynthetic component is based on 
complex indirect arguments and on theoretical models with their natural 
limitations. Such limitations are associated with numerous simplifications
that still enter the modeling of supernovae as well as compact binary mergers, 
with constraints set by the finite numerical resolution and by the omission of
physics that could play a role like, for example, acoustic waves due to 
NS vibration~\cite{qian96,roberts10}, fallback in supernovae and 
associated re-ejection of matter \citep{fryer06}, strong neutron-star magnetic
fields \citep[e.g.][]{thompson03,suzuki05,metzger07,vlasov14},
a rigorous treatment of neutrino oscillations, or non-standard weak-interaction 
physics.

With the goal to assist future experimental searches we provide here 
state-of-the-art estimates of the possible production of the actinides.
In order to circumvent the uncertainties of current core-collapse 
supernova and neutrino-wind models, we adopt an optimistic point of view here
and assume that the r-process taking place in type-II supernovae is capable of 
producing elements up to the heaviest actinides and in addition that each such 
event leads to an r-abundance distribution similar to the one found in the 
solar system. The striking similarity between the solar distribution of r-element
abundances in the $56 \le Z \le 76$ range and the corresponding abundance pattern 
observed in ultra-metal-poor stars like CS 22892-052 \citep{sneden03,sneden08,sneden09} 
led to the conclusion that every astrophysical event producing r-elements gives rise 
to a solar system-like r-abundance distribution, at least for elements above Ba.
Such observations therefore tend to lend support to our assumption, although recent
observations also indicate that star-to-star variations in the r-process content of
metal-poor globular clusters may be a common, although not
ubiquitous, phenomenon \citep{honda07,roederer10,roederer11}.

Section \ref{sect_nass} describes the analytical $\nu$-driven wind model considered
in the present work, while our fitting procedure used to construct an optimal 
reproduction of the solar system r-abundance distribution is outlined in
Sect.~\ref{sect_fit}. The different nuclear inputs used in the present study are
detailed in Sect.~\ref{sect_nuc}. Finally, our fit to the solar system r-abundances, 
the astrophysical ranges required for this fit as well as our predictions concerning
the actinide production are presented in Sect.~\ref{sect_res}. Conclusions are
drawn in Sect.~\ref{sect_conc}.

\section{The wind model}
\label{sect_nass}
Several wind models of analytical or semi-analytical
nature exist. They differ in their level of physical
 sophistication and in their way to parametrize the wind characteristics. In all 
cases, the wind is assumed to be spherically symmetric, which appears to be a 
reasonable first approximation even in two-dimensional simulations, at least at
 late enough times after core bounce \citep{pruet05,arcones11}. 
In addition, the wind is generally treated as a 
stationary flow, meaning no explicit time dependence of any physical  quantity at a
 given radial position $r$, so that $\partial x/\partial t = 0$, let $x$ be the velocity,
 temperature, density, internal energy, pressure, entropy, or composition.
The validity of this approximation is discussed in \cite{thompson01}, where
 it is concluded that stationarity may be reasonably assured, even if some 
caution is warranted. Newtonian, post-Newtonian, and relativistic descriptions of 
spherically symmetric, stationary neutrino-driven winds originating from the surface 
of proto-neutron stars (PNS) have been developed along the lines of a long experience
with previous mathematical treatments of the solar wind and of accretion flows onto 
black holes.

The wind model adopted here corresponds to the analytical Newtonian, adiabatic and 
steady-state wind model, referred to in the following as NASS, derived by \cite{takahashi97}. 
It provides a simple, fully analytical description of the dynamics of the wind outflow at 
relatively late times or sufficiently far away from the PNS surface. The NASS wind 
model relies on the general assumptions listed above, in particular those of a 
Newtonian PNS gravitational potential and of an adiabatic expansion. In addition,  
all the elementary $\nu/\bar{\nu}$ and e$^-$/e$^+$ weak interaction processes are 
assumed to be frozen out, nuclear $\beta$-decays not to affect $Y_{\rm e}$ or $s$,
and possible deviations from nuclear equilibrium with regard to strong and 
electromagnetic interactions to have no influence on the thermodynamical properties 
of the wind. Under such simplifying assumptions, the NASS model cannot predict any 
time variations of the entropy $s$, electron fraction $Y_{\mathrm e}$, or of the 
wind mass loss rate ${\mathrm d}M/{\mathrm d}t \equiv \dot M$, which are thus 
treated as constant input parameters.
 
The  basic NASS wind dynamics in the regime under consideration, and in particular 
for high enough entropies, is well approximated \citep{takahashi97,arnould07}  by 
\begin{equation}
\frac{1}{2}v^2 - \frac{G M_*}{r} + N_{\rm A} k T s_{\rm rad} = E,
\label{eq:wind_motion}
\end{equation}
%
where $M_*$ is the PNS mass, $r$ the radius, $v$ the velocity, $T$ the temperature. 
The total energy per unit mass $E$ may be obtained by setting a boundary condition 
and is usually expressed as a function of the wind energy: $E=f_w \times E_{\rm wind}$, 
where $ E_{\rm wind} = 3 v_s^2/2$ and $v_s$ equals the local adiabatic sound speed 
\citep{arnould07}. For $f_w=1$, the solutions correspond to a sonic wind, whereas for 
$f_w >1 $ they are of subsonic wind type and usually referred to as breeze solutions. 
More details about the wind model can be found in \citet{arnould07}.

The entropy is dominated by photons, electrons and positrons in the wind so that the 
so-called `radiation entropy' $s_{\rm rad}=s_\gamma + s_{e^-} + s_{e^+}$ is given by
\begin{equation}
s_{\rm rad} = s_{\rm rad}^{0} \Bigl[ \frac{4}{11} + \frac{7}{11} 
f_{\rm e} \Bigr]\ \ \ {\rm with}\ \ \ s_{\rm rad}^{0} =
 \frac{11\pi^2}{45 \rho N_{\rm A}} \Bigl( \frac{kT}{\hbar c}  \Bigr) ^3, 
\label{eq:wind_srad}
\end{equation}
%
where $f_{\rm e} = 1$ in the  high-$T$ limit when electrons and positrons are highly 
relativistic, and decreases with $T$ for high $s_{\rm rad}^{0}$-values. In the 
present study, it is approximated by $f_{\rm e}= T_9^2 / (T_9^2 +5.04)$, where $T_9$ 
is the temperature expressed in $10^9$K \citep{witti94}. Since the nucleosynthesis 
in the wind is followed by starting at a temperature of $T_9=9$, the total 
entropy is essentially equal to the radiation entropy (within a few percents). 
Note that the `radiation entropy' is measured here per baryon in units of the 
Boltzmann constant $k_B$ and  includes both the photon and lepton contributions, but 
not the baryonic ones. For the initial conditions adopted here, the latter amounts to 
typically 9~$k_B$/baryon.
 
In the wind model of \cite{takahashi97}, the radial evolution of the velocity, $v(r)$,
and density, $\rho(r)$,
for specified total energy ($E$) per unit mass of outflow material (or,
equivalent, for given $f_w$), given entropy and given mass-loss rate ($\dot M = dM/dt$)
is determined by the energy equation (based on Bernoulli's equation, see
Eq.~\ref{eq:wind_motion}) and the continuity equation $\dot M = 4\pi r^2\rho(r)v(r)$.
The temperature then follows from the assumption of constant entropy, which couples
temperature and density. Varying $f_w$ (or $E$), $s_\mathrm{rad}$ or $\dot M$ therefore
also implies different values of the expansion time scale.
Within the NASS wind model, the velocity at small radii (and high temperatures)
is found to vary as $v\propto \dot M s_{\rm rad}^4$ and consequently
significantly increases with the radiative entropy for given $\dot M$ and $f_w$.

The NASS equations describe the outflow dynamics (wind and breeze
solutions) as functions of radius and as determined by $\dot M$, $E$ (or, equivalent, 
$f_w$), and $s_\mathrm{rad}$. These parameters as well as the electron fraction, $Y_e$,
are considered to be independent of each other, which implies that the NASS model
does not invoke any physical mechanism by which these characteristic parameters 
of the flow are coupled. In reality, the physical processes that drive the mass
outflow from the PNS surface will determine the values of these quantities and may
lead to relations between them. 

The NASS model therefore provides a more general description of outflows from
new-born NSs than, e.g., the usual neutrino-driven wind models mentioned above, 
where neutrino heating near the PNS surface sets the conditions in the ejecta. 
This more general modeling approach is an advantage and requirement
of our study, which intends to explore also the uncertainty limits of
predictions of actinide nucleosynthesis associated with still unsettled aspects of 
the wind physics (like, e.g., the impact of acoustic waves from NS vibrations or strong 
magnetic fields). Because of this freedom of the NASS model, however, we have to
define suitable ranges of the parameters of the model, which allow us to scan the 
range of possibilities. Naturally, this can happen only by the investigation of
a limited number of cases, which we define as ``Ranges I--IV''. 

The four ranges of outflow conditions considered in the present study
are defined as follows, aiming at a good reproduction of the solar r-process
abundances from the first to the third abundance peak. 
In Range~I, we include 150 trajectories with
entropies $50 \le s_{\rm rad}\le 250$ binned in steps of 50, electron fractions 
$0.30 \le Y_e \le 0.48$ in steps of 0.02, mass outflow rates $\dot M = 0.03$, 0.06 and 
$0.30 \times 10^{-5} M_{\odot}/$s, and $f_w=3$.
In Range~II, a set of breeze solutions
is considered, i.e. entropies $50 \le s_{\rm rad}\le 200$
in steps of 50, electron fractions $0.30 \le Y_e \le 0.48$ in steps of 0.02, a single
outflow rate of $\dot M = 0.06 \times 10^{-5} M_{\odot}/$s, and $f_w=3$.
 This  restricted set of 40 trajectories will be shown in Sect.~\ref{sect_res} to 
still allow for a good fit to the solar 
system r-abundances and will consequently be used for a relatively detailed 
sensitivity analysis concerning nuclear uncertainties.

In Range~III, wind solutions are considered, i.e. $f_w=1$, with entropies 
$50 \le s_{\rm rad}\le 250$ in steps of 50, electron fractions $0.30 \le Y_e \le 0.48$ 
in steps of 0.02, and an outflow rate $\dot M = 0.06 \times 10^{-5} M_{\odot}/$s. 

The astrophysical conditions thus considered are associated with a wide range of expansion 
time scales  $\tau_{\rm exp}$.
If we define $\tau_{\rm exp}$ as the time required for the temperature to drop from 
$T_9=9$ to $T_9=2$, $\tau_{\rm exp}$ ranges between 120 and 950~ms for an entropy 
of $s_{\rm rad}=100$ and between 18 and 30~ms for $s_{\rm rad}=250$ at conditions
of Range~I. In Range II (III), $\tau_{\rm exp}=38$ (31)~ms for the largest 
entropies of $s_{\rm rad}=200$, but increases up to 7.3 (7.2)~s for $s_{\rm rad}=50$.

In contrast to the usual behavior of neutrino-driven winds, where higher entropies
are associated with increasing expansion time scales \citep{qian96}, the conditions 
defined by Ranges I--III exhibit inverse relations, namely greater entropies for
shorter expansion time scales. This appeals to some unknown mechanism that can provide
such conditions (which is not implausible in view of the fact that ordinary 
neutrino-driven winds are not found to produce heavy r-process material).
The standard behavior of neutrino-driven winds can be reproduced by our NASS model
by making use of the relation $v\propto \dot M s_\mathrm{rad}^4$ and by modifying the 
outflow rate such that the effect of higher entropies on the wind velocity is
compensated. In order to study a case more compatible with neutrino-driven winds
and breezes, we therefore consider an additional set of breeze solutions, Range~IV,
with the following entropies $s_{\rm rad}$ and outflow rates 
$\dot M$ (in $10^{-5} M_{\odot}/$s): (100, 0.6), (125, 0.17), (150, 0.06), (175, 0.015), 
and (200, 0.006). 
With $f_w=3$, the corresponding expansion time scales $\tau_{\rm exp}$ of these five events 
are 72, 86, 106, 209, and 296~ms, respectively, which clearly increase for increasing 
entropies. For each of these combinations of entropy and outflow rate, 15 events for
electron fractions varying within $0.20 \le Y_e \le 0.48$ in steps of 0.02 are considered.
Range~IV therefore includes 75 events in total.

\section{Fitting the solar system r-abundance distribution}
\label{sect_fit}

In a similar way to that developed in the multi-event canonical s- or r-process models 
\citep{goriely96,goriely97}, it is possible to define a superposition of a large number 
of $\nu$-driven wind components (``events'') that correspond to 
different thermodynamic conditions. Each event is characterized by a given
entropy $s_{\rm rad}$, electron fraction $Y_e$, outflow rate ${\dot M}$, and wind energy 
scaling factor $f_w$.

The combination of r-process events that provides the best fit to the solar abundances 
can then be derived with the aid of an iterative inversion procedure that has been applied
to astronomical inverse problems \citep{lucy74} but also to parametric r-process
calculations \citep{bouquelle96,goriely96,goriely97}. The solar abundance
$N^{\sun}_{Z,A}$ of a nuclide $(Z,A)$ is approximated by the weighted superposition
of the abundances $n(Z,A;s_{\rm rad},Y_e,\dot M,f_w)$ resulting from all astrophysical 
events according to
\begin{equation}
N^{\sun}_{Z,A} \simeq \sum_{s_{\rm rad},Y_e,\dot M,f_w}n(Z,A;s_{\rm rad},Y_e,\dot M,f_w)~\Phi(s_{\rm rad},Y_e,\dot M,f_w)~,
\end{equation}

\noindent where $\Phi(s_{\rm rad},Y_e,\dot M,f_w)$ represents the statistical weight of the event
$(s_{\rm rad},Y_e,\dot M,f_w)$. The recursion relation 
\begin{eqnarray}
\Phi^{(r+1)} (s_{\rm rad},Y_e,\dot M,f_w) && =
\Phi^{(r)}(s_{\rm rad},Y_e,\dot M,f_w)
\sum_{Z,A} \frac{N^{\sun}_{Z,A}}{N_{Z,A}^{(r)}} \cr 
& &\times~ n(Z,A;s_{\rm rad},Y_e,\dot M,f_w)
\end{eqnarray}

\noindent is used in order to obtain an ``improved'' $(r+1)$th estimate of
$\Phi(s_{\rm rad},Y_e,\dot M,f_w)$ from the $r$th iteration
$\Phi^{(r)}$. $N_{Z,A}^{(r)}$ is defined by

\begin{eqnarray}
N_{Z,A}^{(r)} &&= \sum_{s_{\rm rad},Y_e,\dot M,f_w}
n(Z,A;s_{\rm rad},Y_e,\dot M,f_w) \times  \nonumber \\
&&~ \Phi^{(r)}(s_{\rm rad},Y_e,\dot M,f_w) ~.
\end{eqnarray}

\noindent The iteration procedure starts with a uniform distribution of initial
weights $\Phi^{(0)} (s_{\rm rad},Y_e,\dot M,f_w)$ (i.e. all events have initially 
the same weight) and converges after several iterations to a ``best-fit'' abundance
curve $N_{Z,A}$. The corresponding weight profile
$\Phi(s_{\rm rad},Y_e,\dot M,f_w)$ allows us to identify the events that contribute
most significantly to the synthesis of each fitted element.

\begin{figure}
\includegraphics[scale=0.3]{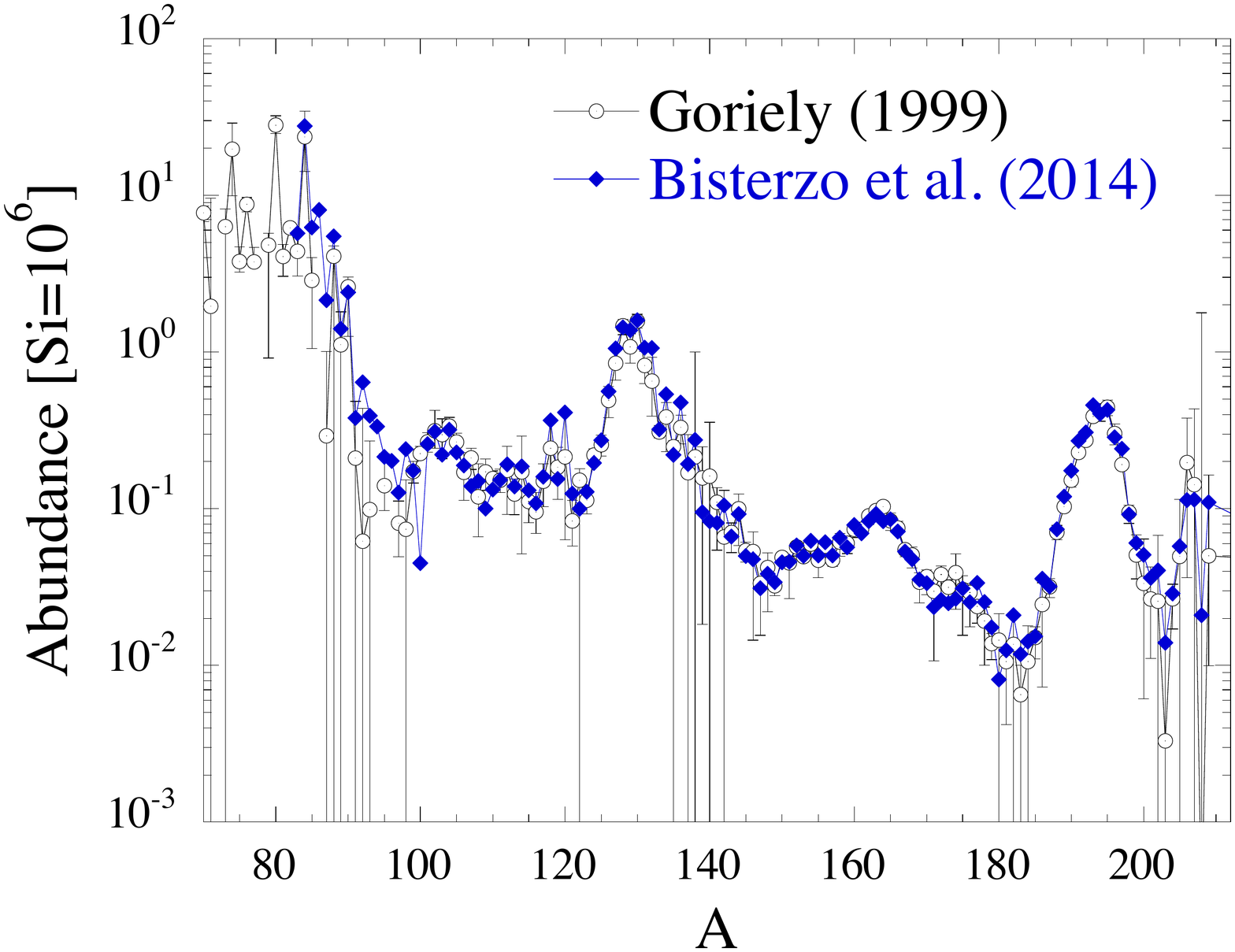}
  \caption{(Color online). Comparison between the r-abundance distributions determined by \citet{goriely99} and \citet{bisterzo14}.}
\label{fig_sol}
\end{figure}

Two different sets of solar system r-abundance distributions are considered in the present study. 
They differ  by the way the s-process contribution to the solar system is estimated. \cite{goriely99} 
used the multi-event canonical s-process model and included a detailed analysis of 
observational, astrophysical and nuclear physics uncertainties, while \cite{bisterzo14} 
estimated the s-process abundances on the basis of a Galactic chemical evolution model 
with  AGB yields based on parametrized $^{13}$C profiles to generate the s-process 
irradiation. Due to the more complex approach, uncertainties are not estimated in the 
latter case and the contribution from the weak s-component in massive stars responsible 
for the solar production of $A \le 90$ nuclei is not included either. The corresponding 
solar r-abundance distributions can differ significantly for s-dominant nuclei, 
especially in the Pb region, as shown in Fig.~\ref{fig_sol}.

\section{Nuclear physics input}
\label{sect_nuc}

The nucleosynthesis is followed with  a reaction network including all 5000 species from protons up 
to $Z=110$ lying  between the valley of $\beta$-stability and the neutron-drip line. All charged-particle 
fusion reactions on light and medium-mass elements that play a role when the nuclear statistical 
equilibrium freezes out are included in addition to radiative neutron captures and photodisintegrations. 
The reaction rates on light species are taken from the NETGEN library, which includes all the latest 
compilations of experimentally determined  reaction rates \citep{xu13}. By default, experimentally unknown reactions 
are estimated  with the TALYS code \citep{goriely08,koning12} on the basis of the HFB-21 nuclear masses \citep{goriely10}, the HFB plus combinatorial nuclear level densities \citep{goriely08b} and the QRPA E1 strength functions \citep{goriely04}. 
Fission and $\beta$-decay processes are also 
included, i.e. neutron-induced fission, spontaneous fission, $\beta$-delayed 
fission, as well as $\beta$-delayed neutron emission \citep{goriely15}. The  
$\beta$-decay processes are taken from the updated version of the 
Gross Theory \citep{tachibana90} based on the HFB-21 $Q$-values \citep{goriely10}, when not available 
experimentally, whereas all fission processes are estimated on the basis of the HFB-14 
fission paths \citep{goriely07} and the full calculation of the corresponding barrier penetration 
\citep{goriely09}. The fission fragment distribution is taken from the SPY model as described 
in \citet{goriely13}. This nuclear physics set represents our standard input. Due to the large uncertainties still affecting the properties of the neutrinos, in particular their luminosities and temperatures, no neutrino interactions on nuclei are included in the present calculations. (Neutrino interactions with free nucleons in the close vicinity of the neutron star are considered to be important in setting the dynamical and thermodynamic conditions and the neutron-to-proton ratio in the wind outflow, which is inherently accounted for by our parametric wind model.)

To estimate the sensitivity of the abundance calculations with respect to the still uncertain nuclear physics predictions far away from the valley of $\beta$-stability, different nuclear ingredients are also considered in addition to the above-described standard set. These include
\begin{itemize}
\item HFB-31: reactions rates determined on the basis of the Hartree-Fock-Bogoliubov (HFB) HFB-31 mass model \citep{goriely16}; all other inputs remaining identical;
\item D1M: reactions rates determined on the basis of the D1M Gogny-HFB mass model \citep{goriely09b}; all other inputs remaining identical;
\item FRDM: reactions rates determined with the 2012 version of the FRDM mass model  \citep{moller16}, the back-shifted Fermi Gas model for nuclear level densities \citep{koning08} and Lorentzian-type $E1$ strength function \citep{kopecky90}; $\beta$-decay rates are taken from the Random Phase Approximation and FRDM-based $Q$-values \citep{moller03}; Fission probabilities are based on the \citet{ms99} fission barriers and the fragment yields on the GEF model \citep{gef12}. This nuclear physics set essentially includes so-called phenomenological macroscopic approaches in contrast to the microscopic input characterizing the standard set; 
\item FIS: reaction and $\beta$-decay rates are identical to the default models, but fission probabilities are based on the \citet{ms99} fission barriers and the fragment yields on the GEF model \citep{gef12};
\item BETA: $\beta$-decay rates from the Gross Theory are replaced by the Tamm-Dancoff approximation~\citep{klapdor84}; all other inputs remaining identical to the standard set.
\end{itemize}

\section{Results}
\label{sect_res}

\subsection{Actinide production}
\begin{figure}
\includegraphics[scale=0.3]{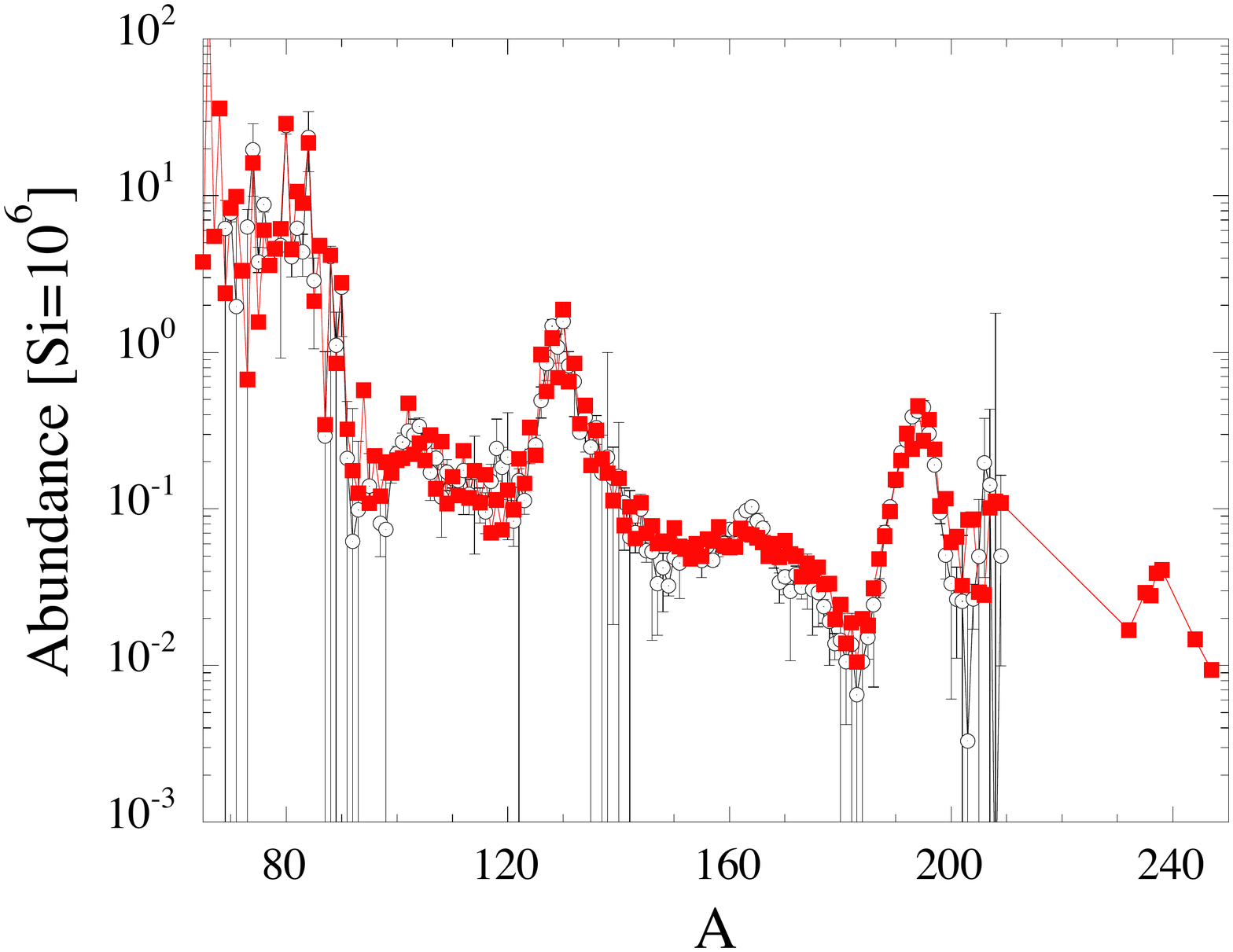}
  \caption{(Color online). Fit to the solar system r-abundances \citep{goriely99} obtained with the multi-event NASS wind model for events taking place with Range~I conditions and the default nuclear physics input.}
\label{fig_afit}
\end{figure}

\begin{figure}
\includegraphics[scale=0.3]{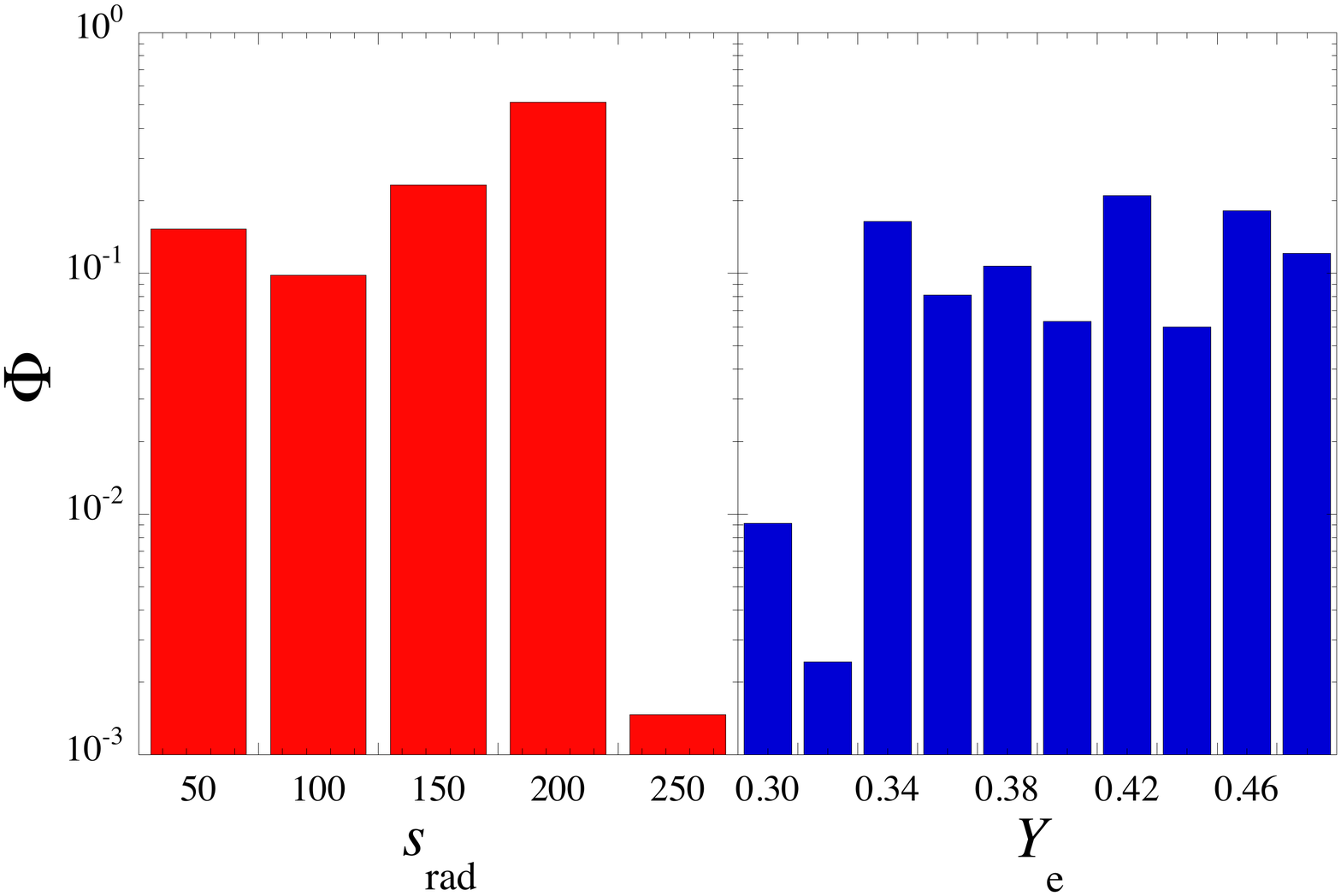}
  \caption{(Color online). 
 Histograms of the statistical weights $\Phi_s=\sum_{Y_e,\dot M,f_w}\Phi(s_{\rm rad},Y_e,\dot M,f_w)$ as a function of the radiative entropy $s_{\rm rad}$ (left panel) and $\Phi_{Y_e}=\sum_{s_{\rm rad},\dot M,f_w}\Phi(s_{\rm rad},Y_e,\dot M,f_w)$ as a function of electron fraction $Y_e$ (right panel) responsible for the abundance fit shown in Fig.~\ref{fig_afit} for Range~I conditions.}
\label{fig_histo}
\end{figure}

\begin{figure}
\includegraphics[scale=0.3]{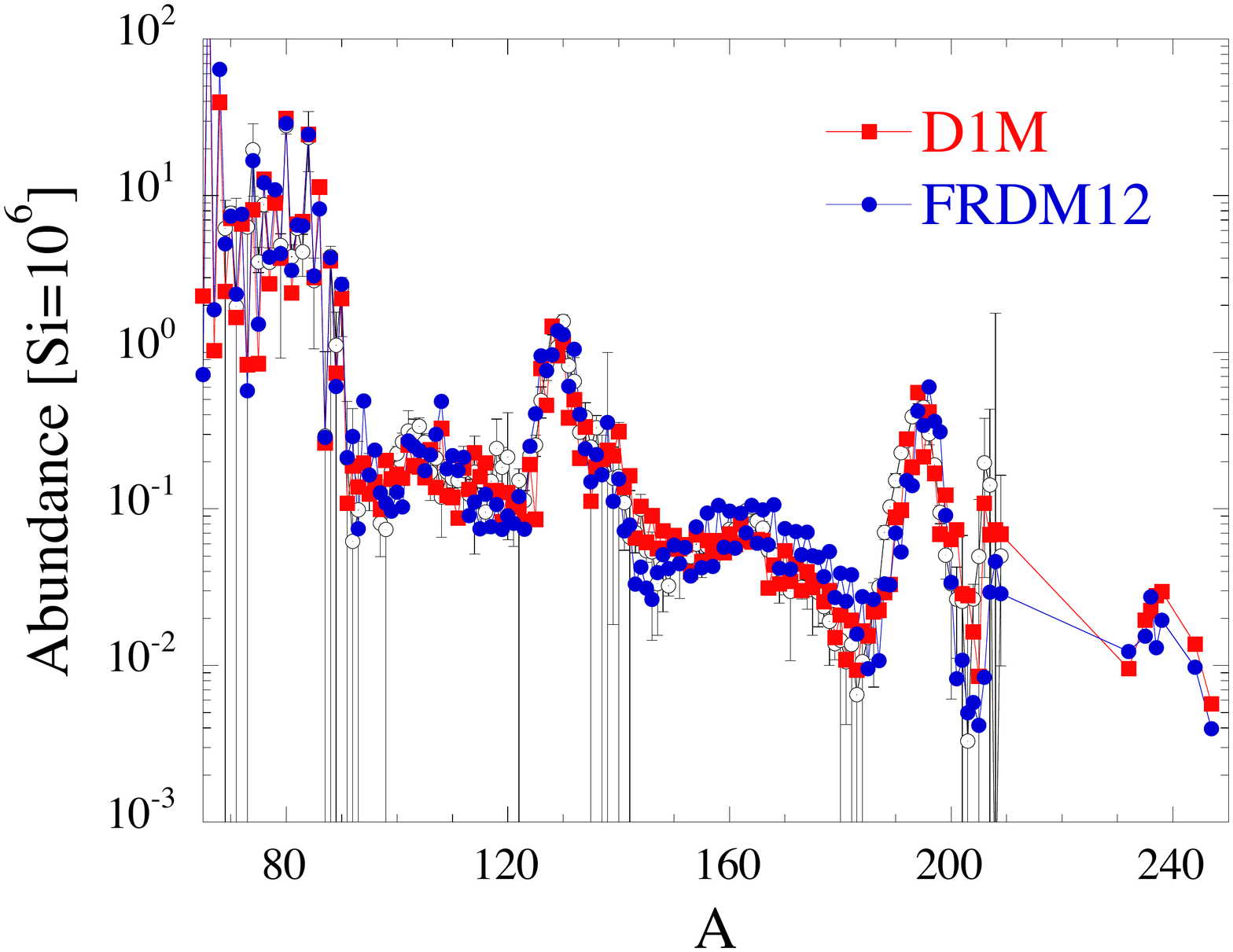}
  \caption{(Color online). Same as Fig.~\ref{fig_afit} but for events 
with Range~II conditions and the nuclear physics inputs corresponding to D1M and FRDM.}
\label{fig_afit2}
\end{figure}

\begin{figure}
\includegraphics[scale=0.3]{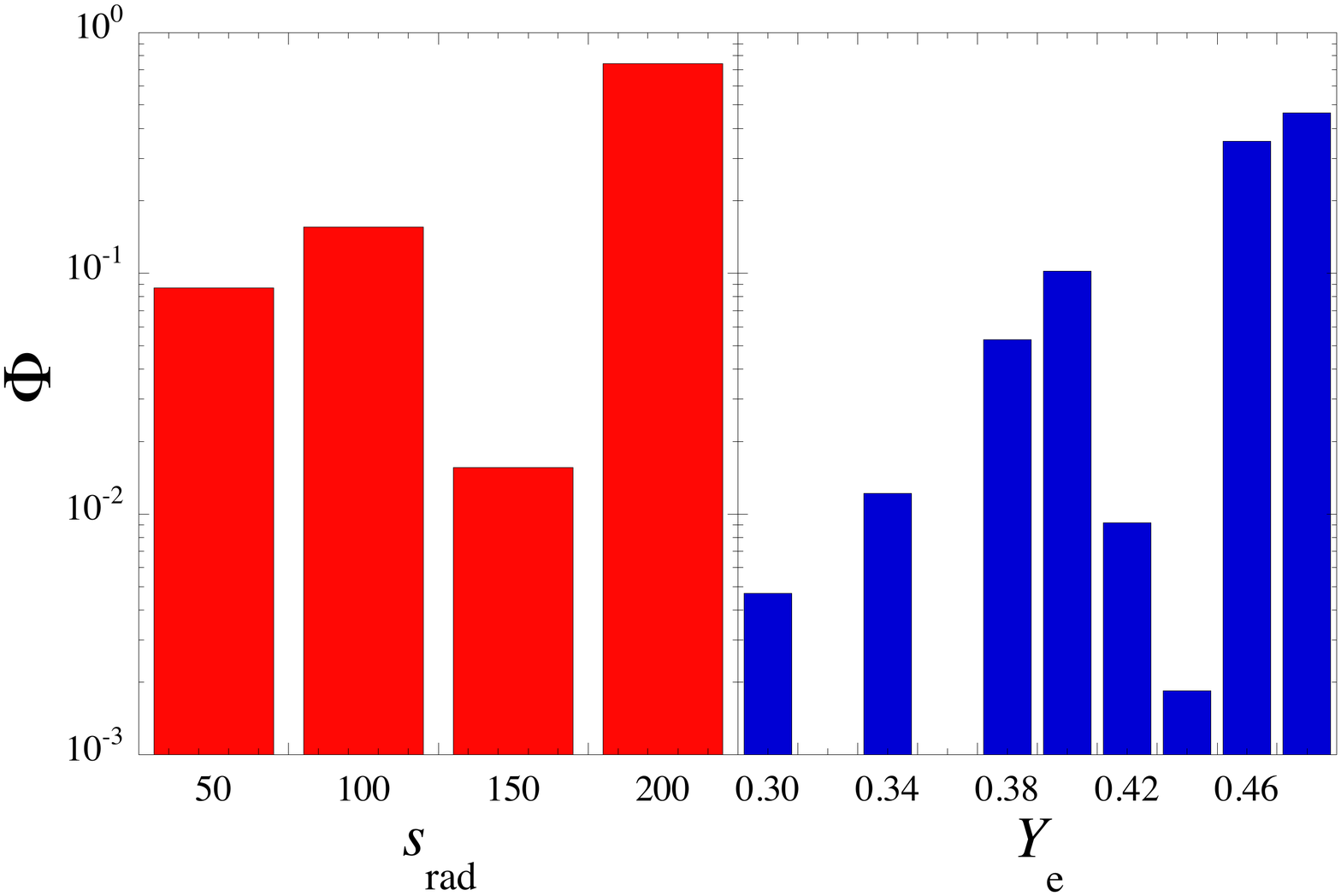}
  \caption{(Color online). 
  Same as Fig.~\ref{fig_histo} but for the statistical weights corresponding to the abundance fit shown in Fig.~\ref{fig_afit2} with D1M mass model and Range~II conditions.}
\label{fig_histo2}
\end{figure}

The solar system abundances have been fitted assuming different ranges of astrophysical conditions for the $\nu$-driven wind and different nuclear physics inputs, as described in the previous sections. One example (Case 1; Table~\ref{tab1}) of a fit corresponding to the Range~I of astrophysical conditions and the standard nuclear physics input, is shown in Fig.~\ref{fig_afit}. It can be seen that the solar distribution \citep[in this case, taken from ][]{goriely99} is rather well fitted, although deviations can be observed, in particular in the rare-earth region. The corresponding distributions of entropies and electron fractions are illustrated for this standard case in Fig.~\ref{fig_histo}. For illustration, additional fits with Range~II of astrophysical conditions and the nuclear physics inputs corresponding to D1M and FRDM (Cases 7 and 11, respectively; Table~\ref{tab1}) are given in Fig.~\ref{fig_afit2}. It will also be noticed that the restricted Range~II is already sufficient to provide the conditions needed to reproduce the solar system distribution fairly well. The statistical weights of entropies and electron fractions needed to obtain these fits are obviously dependent on the adopted range of astrophysical conditions as well as the nuclear physics inputs, as shown in Fig.~\ref{fig_histo2}.

Considering different combinations for the solar r-abundance distributions, ranges of thermodynamic conditions and nuclear physics inputs, quite some different predictions for the production of actinides can be derived as summarized in Table~\ref{tab1}. In particular, it can be seen that wind solutions ($f_w=1$) in the Range~III of thermodynamic conditions lead to a significantly lower production of some actinides compared to the breeze solutions ($f_w=3$ in Range~II). The specific conditions defined in Range~IV lead to an actinide production quite 
similar to the one obtained in Range~II.  

Nuclear physics inputs, in particular nuclear mass models (and the corresponding rates) as well as $\beta$-decay rates, have also a non-negligible impact on the abundance predictions, as already well established for decades. Uncorrelated lower and upper limits to the production of the long-lived actinides can be deduced and are also given in Table~\ref{tab1}. The uncertainties for the $^{232}$Th production are found to be considerably larger than those for the production of $^{244}$Pu and $^{247}$Cm. For comparison, the upper and lower limits estimated within the multi-event canonical model \citep{goriely01} on the basis on very different astrophysical conditions and nuclear physics inputs are also given in Table~\ref{tab1}.

For completeness, Table~\ref{tab1} also gives the estimated age of the ultra-metal-poor star CS 31082-001. Indeed, accurate observations of heavy r-elements have been used to estimate the age of ultra-metal-poor stars on the basis of the fundamental assumption that the r-process is universal \citep{sneden96,cowan97,goriely99b}. In particular,  if Th and U lines could be observed accurately and simultaneously in metal-poor stars, a relatively reliable age estimate could be derived from the expression
\begin{equation}
\log \Bigl({{\rm Th} \over {\rm U}} \Bigr)_{\rm obs} = \log \Bigl({{\rm Th} \over {\rm U}}
\Bigr)_{r} +\log {\rm e} ~ \Bigl( {1 \over \tau({\rm U})} - {1\over \tau({\rm Th})}\Bigr)~T^*_{\rm{U,Th}} ,
\label{eq2}
\end{equation}
\noindent where $\tau({\rm U})=6.45~{\rm Gyr}$ is the characteristic $\alpha$-decay time\-scale of $^{238}$U and similarly $\tau({\rm Th})=20.27~{\rm Gyr}$ for $^{232}$Th. Note that the Th abundance in the r-process ratio of Eq.~(\ref{eq2}) corresponds to the $^{232}$Th value after decay of its shorter-lived progenitors $^{236}$U  and  $^{244}$Pu, and the U abundance corresponds to the $^{238}$U contribution only. Th and U lines have been observed in particular in the CS 31082-001 star with a $\log ({\rm U/Th})=-0.94 \pm 0.09$ \citep[][and private communication]{cayrel01}. The 0.09 dex observational error gives rise to a $\pm 2$~Gyr uncertainty on the estimate of $T^*_{\rm{U,Th}}$. For each simulation, the estimated age of CS 31082-001  is given in Table~\ref{tab1}. The ages obtained are reasonably consistent with the WMAP estimate of 13.8~Gyr \citep{planck15}.

\begin{table*}
  \caption{Abundances (normalized to Si=$10^6$) of Pb and of the actinides with
half-lives $t_{1/2} > 10^6$ y predicted by the multi-event NASS wind model with solar system r-abundances  SOL1 \citep{goriely99} or SOL2 \citep{bisterzo14} and for astrophysical conditions constrained to the Ranges I, II, III or IV (see Sect.~\ref{sect_nass}). The last column gives the $T^*_{\rm{U,Th}}$ age (in Gyr) of CS 31082-001 based on the U/Th cosmochronometry. The calculations are also based on the various
indicated combinations of nuclear inputs concerning masses, fission, $\beta$-decay and reaction
rates (see Sect.~\ref{sect_nuc} for more details). The three lines (Rec, Min, Max) correspond to the recommended abundances with an estimate of
the minimum and maximum values based on the above calculations, while the last two lines give, for comparison, the upper and lower limits estimated within the multi-event canonical model \citep[][GA01]{goriely01}.}

   \begin{tabular}{|c|ccc|ccccccccc|}
  \hline
Case	&	Range 	& SOL &	Nuc	&	Pb	&	$^{232}$Th	&	$^{235}$U	&	$^{236}$U	&	$^{238}$U	& $^{237}$Np & $^{244}$Pu 	&	$^{247}$Cm	&  $T^*_{\rm{U,Th}}$ \\
\hline																					
1&	I	&	1	&	Std	&	2.53E-01	&	2.45E-02	&	4.27E-02	&	4.12E-02	&	5.66E-02	&	5.48E-02	&	1.87E-02	&	1.24E-02	&	16.69	\\
2&	II	&	1	&	Std	&	2.48E-01	&	2.03E-02	&	3.50E-02	&	3.38E-02	&	4.44E-02	&	4.39E-02	&	1.43E-02	&	9.89E-03	&	16.37	\\
3&	III	&	1	&	Std	&	1.15E-01	&	8.07E-04	&	2.60E-03	&	1.52E-03	&	2.53E-03	&	4.99E-03	&	3.35E-03	&	3.09E-03	&	12.83	\\
4&	IV	&	1	&	Std	&	2.51E-01 	&  	1.86E-02 	&  	3.12E-02 	&  	2.73E-02 	& 	 3.88E-02 &	 3.79E-02 &  	1.37E-02 	&  	7.49E-03 	&   	16.40		\\
5&	I	&	2	&	Std	&	1.98E-01	&	1.96E-02	&	3.46E-02	&	3.28E-02	&	4.70E-02	&	4.48E-02	&	1.61E-02	&	1.00E-02	&	16.90		\\
6&	II	&	2	&	Std	&	2.08E-01	&	1.69E-02	&	2.95E-02	&	2.89E-02	&	4.02E-02	&	3.92E-02	&	1.35E-02	&	9.19E-03	&	16.78	\\
7&	III	&	2	&	Std	&	9.92E-02	&	1.22E-03	&	3.91E-03	&	2.05E-03	&	3.89E-03	&	8.03E-03	&	5.33E-03	&	5.17E-03	&	12.97	\\
8&	IV	&	2	&	Std	&	2.07E-01  &  	1.52E-02 	&  	2.64E-02 	&  	2.21E-02 	&  	3.64E-02 	&  	3.38E-02 	&  	1.30E-02 	&  	6.48E-03 	&   	17.39	\\
9&	II	&	1	&	HFB-31	&	3.63E-01 &  3.41E-02 &  	7.71E-02 &  7.33E-02 &  8.53E-02 &  9.29E-02 &  3.89E-02 &  1.95E-02 &   15.36	\\
10&	II	&	2	&	HFB-31	&	3.21E-01 &  2.95E-02 &  6.57E-02 &  6.18E-02 &  7.48E-02 &  8.04E-02 &  3.66E-02 &  1.72E-02 &   15.38	\\
11&	II	&	1	&	FRDM	&	8.73E-02	&	1.28E-02	&	1.60E-02	&	2.86E-02	&	2.02E-02	&	1.36E-02	&	1.02E-02	&	4.11E-03	&	11.62	\\
12&	III	&	1	&	FRDM	&	1.65E-01	&	3.05E-03	&	8.27E-03	&	1.03E-02	&	1.37E-02	&	9.47E-03	&	6.15E-03	&	3.27E-03	&	17.12	\\
13&	II	&	2	&	FRDM	&	8.75E-02	&	1.27E-02	&	1.46E-02	&	2.51E-02	&	1.78E-02	&	1.20E-02	&	9.14E-03	&	3.67E-03	&	11.29	\\
14&	III	&	2	&	FRDM	&	1.59E-01	&	2.90E-03	&	7.79E-03	&	9.74E-03	&	1.30E-02	&	8.91E-03	&	5.79E-03	&	3.06E-03	&	17.16	\\
15&	II	&	1	&	D1M	&	2.88E-01	&	1.10E-02	&	2.25E-02	&	2.57E-02	&	3.41E-02	&	3.20E-02	&	1.57E-02	&	6.55E-03	&	16.39	\\
16&	II	&	2	&	D1M	&	2.33E-01	&	8.25E-02	&	1.65E-02	&	1.87E-02	&	2.77E-02	&	2.39E-02	&	1.28E-02	&	4.51E-03	&	17.04	\\
17&	II	&	1	&	FIS	&	2.28E-01	&	1.90E-02	&	3.18E-02	&	3.44E-02	&	3.80E-02	&	3.65E-02	&	1.07E-02	&	9.82E-03	&	15.51	\\
18&	II	&	2	&	FIS	&	1.97E-01	&	1.79E-02	&	3.11E-02	&	3.50E-02	&	4.20E-02	&	3.80E-02	&	1.28E-02	&	1.20E-02	&	16.22	\\
19&	II	&	1	&	BETA	&	2.48E-01	&	2.35E-02	&	3.09E-02	&	3.04E-02	&	2.74E-02	&	2.99E-02	&	1.10E-02	&	6.70E-03	&	12.31	\\
20&	II	&	2	&	BETA	&	2.48E-01	&	2.45E-02	&	3.23E-02	&	3.32E-02	&	3.06E-02	&	3.27E-02	&	1.26E-02	&	7.62E-03	&	12.61	\\
\hline
Rec	&	I	&	1	&	Std	&	2.53E-01	&	2.45E-02	&	4.27E-02	&	4.12E-02	&	5.66E-02	&	5.48E-02	&	1.87E-02	&	1.24E-02	&	16.69	\\							
Min		&		&		&	 &	8.73E-02	&	8.07E-04	&	2.60E-03	&	1.52E-03	&	2.53E-03	&	4.99E-03	&	3.35E-03	&	3.06E-03	&	11.29	\\
Max		&		&		&	 &3.63E-01	&	8.25E-02	&	7.71E-02	&	7.33E-02	&	8.53E-02	&	9.29E-02	&	3.89E-02	&	1.95E-02	&	17.16	\\
\hline
	Min$^*$	&   GA01	&		&	         	&	5.09E-01	&	2.53E-02	&	2.26E-02	&	2.15E-02	&	2.32E-02	&	1.46E-02	&	3.19E-03	&	2.00E-03	& 8.94 \\
	Max$^*$	&   GA01       	&		&		        &	8.69E-01	&	6.77E-02	&	1.13E-01	&	1.00E-01	&	1.77E-01	&	1.03E-01	&	1.46E-01	&	3.62E-02	& 17.73 \\
				
\hline																		
\end{tabular}
\label{tab1}
\end{table*}

%
%

\subsection{Actinide yields per supernova event}

Table~\ref{tab1} gives the abundances of Pb and of the main actinides (in the Si=$10^6$ scale) produced by the r-process in the $\nu$-driven wind, assuming the r-process is universal and produces an abundance distribution identical to the solar one. To estimate the yields of ejected material per supernova event, we have to know the total amount of matter accumulated in the $\nu$-driven wind and able to make r-process elements. Since no realistic $\nu$-driven wind model exists that leads to a successful r-process, this quantity remains unknown. For this reason, we adopt here a fiducial value of $M_{\rm wind}=7 \times 10^{-4}$\Msun\ ejected per Galactic supernova as well as an amount of the r-only nucleus $^{130}$Te produced by each supernova of about $6 \times 10^{-6}$\Msun\  \citep{taka94}. Using now an abundance of 1.59 of $^{130}$Te in the Si=$10^6$ scale (Figs.~\ref{fig_sol}, \ref{fig_afit}, \ref{fig_afit2}), it is straightforward to estimate the ejected yields for each actinide. The upper and lower limits of these yields are given in Table~\ref{tab2} on the basis of the minimum and maximum abundances given in Table~\ref{tab1}. Clearly, a large uncertainty factor should also be applied to the $M_{\rm wind}$ value and correspondingly to the yields, or equivalently the yields of Table~\ref{tab2} should be taken proportional to the $M_{\rm ej}/(7 \times 10^{-4}$\Msun) where $M_{\rm ej}$ is the still unknown total mass of wind ejecta per supernova that contribute to the r-process production. A rather firm theoretical upper limit for this number may be a few $10^{-3}\,M_\odot$ (close to the maximum total mass of neutrino-driven wind ejecta; R. Bollig, private communication), but according to present state-of-the-art supernova models $M_{\rm ej}$ is expected to vanish \citep{hudepohl10,fisch10,janka12,mirizzi16}. It should also be recalled here that our numbers in Table~\ref{tab2} rely on the fundamental assumption that the r-process taking place in core-collapse supernovae is capable of producing elements up to the heaviest actinides and, in addition, that each of such events leads to an r-abundance distribution similar to the one found in the solar system.


\begin{table}
  \centering
  \caption{Upper and lower limits of the yields (in \Msun) of Pb and  long-lived actinides ejected 
per supernova assuming the matter accumulated in the $\nu$-driven wind corresponds to 
$7 \times 10^{-4}$\Msun\ \citep{taka94}.}

   \begin{tabular}{|c|cc|}
  \hline
  Nuc & Min & Max \\
  \hline
    Pb			& 5.26E-07 &	2.19E-06 \\
$^{232}$Th	& 5.42E-09 &	5.54E-07 \\
$^{235}$U		& 1.77E-08 &	5.25E-07 \\
$^{236}$U		& 1.04E-08 &	5.01E-07 \\
$^{238}$U		& 1.74E-08 & 	5.88E-07 \\
$^{237}$Np	& 3.42E-08 & 	6.38E-07 \\
$^{244}$Pu	& 2.37E-08 &	2.75E-07 \\
$^{247}$Cm	& 2.19E-08 &	1.39E-07 \\
\hline																	
\end{tabular}
\label{tab2}
\end{table}

\section{Conclusions}
\label{sect_conc}

We derived updated estimates of the production of the radioactive  
actinides of $^{232}$Th, $^{235,236,238}$U, $^{237}$Np,
$^{244}$Pu, and $^{247}$Cm and of the corresponding uncertainty
ranges based on the assumption that the neutrino-driven wind of
the proto-neutron star in core collapse supernovae is able to
provide r-process viable conditions and can well reproduce the 
universal r-process abundance pattern observed in the Sun and 
found in metal-poor stars. Since current state-of-the-art 
hydrodynamical models of supernova explosions and of the 
neutrino-driven wind do not yield the dynamic and thermodynamic
conditions needed for the production of heavy r-process material,
we base our study on the simple, analytical NASS (Newtonian,
adiabatic, steady-state) wind model and take into account the 
mass-weighted superposition of a large number of wind 
components with different nucleosynthesis-relevant characteristics
of entropy, expansion time scale, and neutron excess. The set of
chosen conditions is constrained by the ability of the integrated
wind material to match the solar r-abundance distribution.
In order to explore the sensitivity of our estimates for the
actinide production to uncertain nuclear physics, we tested the
influence of six different sets of nuclear-rate ingredients.

The consideration of neutrino-driven winds of supernovae as
possible r-process site, despite the unfavorable conditions
provided by current hydrodynamical models, and the combination
of wind components under the mentioned constraint, can be 
justified by the fact that also the state-of-the-art models 
still suffer from major uncertainties, e.g.\ with respect to
neutrino opacities in correlated nuclear matter, the effects
of potentially strong magnetic fields inside and around the
nascent NS, or the incomplete exploration of 
neutrino-oscillation effects and possible non-standard weak
interaction physics in the PNS environment.
On the other hand, despite considerable modeling progress and
generally more promising properties of the ejecta dynamics,
also NS-NS and/or NS-BH mergers are by far not established as
the main sources of heavy r-process nuclei, because  in
this case too, major uncertainties still affect the weak-interaction sector
as well as the description of strong magnetic field effects.

Further observational efforts for a positive confirmation
of the one or the other or of both possible sources are therefore
needed. This includes searches of electromagnetic transients 
associated with the radiation emission of NS-NS/BH merger ejecta 
heated by the radioactive decay of r-process nuclei 
\citep[see for example,][]{metzger10,roberts11,goriely11,korobkin12,bauswein13,kasen15,martin15}
as well as the exploration of cosmic and terrestrial reservoirs
for signatures of freshly produced r-nuclei. Our work is
intended to facilitate the observational and experimental
analyses of the latter kind and their interpretation.

We find a considerable spread of the results of actinide yields
with the largest uncertainty (a factor of $\sim$100) for $^{232}$Th
and the lowest uncertainty for $^{244}$Pu (factor of $\sim$12) 
and for $^{247}$Cm (factor of $\sim$7), with supersonically expanding 
wind solutions producing considerably less actinide material than
more slowly, subsonically expanding breeze outflows. 
For investigations of terrestrial material like recently 
performed by \cite{wallner15}, it is interesting to know the
isotope-production ratio of $^{244}$Pu compared to the long-lived
radioactive nucleus $^{60}$Fe. Current supernova models
predict a yield of $^{60}$Fe of $\sim 3\times 10^{-5}\,M_\odot$
per massive-star death or $\sim 4\times 10^{-5}\,M_\odot$ per supernova
\citep{sukhbold15}. Since this theoretical value seems to exceed
estimates based on cosmic-rays near the Earth by about a factor
of 2 \citep{sukhbold15}, we consider here a $^{60}$Fe output per
exploding massive star of $\sim 2\times 10^{-5}\,M_\odot$. 
With this number we obtain a theoretical range for the
supernova-produced $^{244}$Pu/$^{60}$Fe isotope ratio between
roughly $1.2\times 10^{-3}$ and $1.4\times 10^{-2}$, using the
yields of Table~\ref{tab2}. With experimentally determined
limits of less than $10^{-4}$ for their crust and sediment samples, 
\cite{wallner15} set a bound still more than a factor of 10 below
our lower limit. Therefore this bound
seems to exclude a recent insemination of the Earth by r-process
material from a frequent source like supernovae, in particular
from a nearby supernova $\sim$2.2\,My in the past. However,
the exclusion may not be as convincing as suggested by the 
two orders of magnitude discrepancy advocated by \cite{wallner15}.
Not only the probability with which $^{244}$Pu from a possible
supernova origin ends up in the investigated material samples
is uncertain, also the extreme sensitivity of the actinide 
production to the model variations tested in our work should
be taken as a warning.

\section*{Acknowledgments}
We thank Robert Bollig for useful discussions and Shawn Bishop 
for inspiring conversations as well as comments on the manuscript.
 We are grateful to the referee, Stan Woosley, for constructive
questions that helped us to improve the presentation.
SG acknowledges financial support from FRS-FNRS (Belgium), HTJ from
Deutsche Forschungsgemeinschaft through the Cluster of Excellence
``Origin and Structure of the Universe'' (EXC-153).

\label{lastpage}

\begin{thebibliography}{}

\bibitem[Ade et al.(2015)]{planck15} Ade P.A.R. et al., 2015, arXiv:1502.01589
\bibitem[Arcones \& Janka(2011)]{arcones11} Arcones A. \& Janka H.-T., 2011, \aap, 526, A160 
\bibitem[Argast et al.(2004)]{argast04} Argast, J. et al., 2004, \aap, 416, 997
\bibitem[Arnould et al.(2007)]{arnould07} Arnould M., Goriely S., Takahashi K.,  2007, \physrep, 450, 97
\bibitem[Banerjee et al.(2011)]{banerjee11} Banerjee P., Haxton W.C., Qian Y.-Z., 2011, \prl, 106, 201104  
\bibitem[Bauswein et al.(2013)]{bauswein13}  Bauswein A., Goriely S.,  Janka H.-T., 2013, \apj, 773, 78
\bibitem[Bishop \& Egli(2011)]{bishop11} Bishop S. \& Egli R., 2011, Icarus,  212, 960
\bibitem[Bisterzo et al.(2014)]{bisterzo14} Bisterzo S., Travaglio C., Gallino R., Wiescher M., K\"appeler F., 2014, \apj,  787, 10
\bibitem[Bouquelle et al.(1996)]{bouquelle96} Bouquelle V., Cerf N., Arnould M., Tachibana T., Goriely S., 1996, \aap, 305, 1005
\bibitem[Cayrel et al.(2001)]{cayrel01} Cayrel R., Hill V., Beers T.C., et al., 2001, Nature, 409, 691
\bibitem[Cowan et al.(1997)]{cowan97} Cowan J.J., McWilliam A., Sneden C., Burris D.L., 1997, \apj, 480,246 
\bibitem[Dominik et al.(2012)]{dom12} Dominik,  M., Belczynski, K., Fryer, C.,  Holz, D.E.,  Berti, E., Bulik, T.,  Mandel, I. \& O'Shaughnessy, R. 2012, \apj, 759, 52
\bibitem[Eichler et al.(1989)]{eichler89} Eichler D., Livio M., Piran T., Schramm D.N., 1989, Nature, 340, 126 
\bibitem[Fischer et al.(2010)]{fisch10} Fischer T., Whitehouse S.C., Mezzacappa A., Thielemann F.-K., Liebend\"orfer  M., 2010,  \aap, 517, A80
\bibitem[Fitoussi et al.(2008)]{fitoussi08} Fitoussi C., Raisbeck G. M., Knie K., Korschinek G., Faestermann  T. et al., 
2008, \prl, 101, 121101
\bibitem[Fowler \& Hoyle(1960)]{fowler60} Fowler W.A. \& Hoyle F., 1960, Ann. Phys., 10, 280
\bibitem[Freiburghaus et al.(1999)]{frei99}  Freiburghaus C.,  Rosswog S., Thielemann F.-K., 1999, \apj, 525,  L121
\bibitem[Fryer et al.(2006)]{fryer06} Fryer C.L., Herwig F., Hungerford A., Timmes F.X., 2006, \apjl, L131
\bibitem[Goriely \& Arnould(1996)]{goriely96} Goriely S., Arnould M., 1996, \aap,  312, 327
\bibitem[Goriely(1997)]{goriely97} Goriely S., 1997, \aap,  327, 845
\bibitem[Goriely(1999)]{goriely99} Goriely S., 1999, \aap,  342, 881
\bibitem[Goriely \& Clerbaux(1999)]{goriely99b} Goriely S., Clerbaux B., 1999b, \aap,  346, 798
\bibitem[Goriely \& Arnould(2001)]{goriely01} Goriely S., Arnould M., 2001, \aap,  379, 1113
\bibitem[Goriely et al.(2004)]{goriely04} Goriely S., Khan E., Samyn M., 2004, Nucl. Phys., A739, 331
\bibitem[Goriely et al.(2007)]{goriely07} Goriely S., Samyn M., Pearson  J. M., 2007, \prc, 75, 064312
\bibitem[Goriely et al.(2008)]{goriely08} Goriely S., Hilaire S., Koning A.J., 2008, \aap, 487, 767 
\bibitem[Goriely et al.(2008b)]{goriely08b} Goriely S., Hilaire S., Koning A.J., 2008b, \prc, 78, 064307
\bibitem[Goriely et al.(2009)]{goriely09} Goriely S., Hilaire S., Koning A.J., Sin M., Capote R., 2009, \prc, 79,  024612
\bibitem[Goriely et al.(2009b)]{goriely09b} Goriely S., Hilaire S., Girod M, P\'eru S., 2009b, \prl, 102,  242501
\bibitem[Goriely et al.(2010)]{goriely10} Goriely S., Chamel N., Pearson  J. M., 2010, \prc, 82, 035804
\bibitem[Goriely et al.(2011)]{goriely11} Goriely S.,  Bauswein A.,  Janka H.-T., 2011, \apjl, 738, L32
\bibitem[Goriely et al.(2013)]{goriely13} Goriely S., Sida J.-L.,  Lema\^\i tre J.-F.,  Panebianco S., Dubray N., Hilaire S., Bauswein A.,  Janka H.-T.,  2013, \prl,  111, 242502
\bibitem[Goriely(2015)]{goriely15} Goriely S., 2015, European Physical Journal, A51, 172
\bibitem[Goriely et al.(2015)]{goriely15a} Goriely S., Bauswein A., Just O., Pllumbi E., Janka H.-Th., 2015, \mnras, 452, 3894
\bibitem[Goriely et al.(2016)]{goriely16} Goriely S., Chamel N., Pearson  J. M., 2016, \prc, 93, 034337 
\bibitem[Hoffman et al.(1997)]{hoffman97} Hoffman R.D., Woosley S.E., Qian Y.-Z., 1997, \apj, 482, 951
\bibitem[Honda et al.(2007)]{honda07} Honda S., Aoki W., Ishimaru Y., Wanajo S., 2007, \apj, 666, 1189
\bibitem[Hotokezaka et al.(2015)]{hotokezaka15} Hotokezaka K., Piran T., Paul M., 2015, Nature Physics, 11, 1042
\bibitem[H\"udepohl et al.(2010)]{hudepohl10} H\"udepohl L., M\"uller B., Janka H.-T., Marek A., Raffelt G.G., 2010, \prl, 104, 251101
\bibitem[Ishimaru et al.(2015)]{ishimaru15} Ishimaru Y., Wanajo S., Prantzos N., 2015, \apj, 804, L35 
\bibitem[Janka(2012)]{janka12} Janka H.-T., 2012, Ann. Rev. Nuc. Part. Science, 62, 407
\bibitem[Just et al.(2015)]{just15} Just O., Bauswein A., Ardevol Pulpillo R., Goriely S., Janka H.-T., 2015, \mnras, 448, 541
\bibitem[Kasen et al.(2015)]{kasen15} Kasen D., Fern{\'a}ndez R. \& Metzger B.D., 2015, \mnras, 50, 1777
\bibitem[Kim et al.(2010)]{kim10} Kim C., Kalogera V. \& Lorimer D.R., 2010, New Astron. Rev., 54, 148
\bibitem[Klapdor et al.(1984)]{klapdor84} Klapdor H. V., Metzinger  J., Oda T., At. Data Nucl. Data Tables, 1984, 31, 81
\bibitem[Knie et al.(1999)]{knie99} Knie K., Korschinek G., Faestermann T., Wallner C., Scholten J., Hillebrandt W., \prl, 83, 18
\bibitem[Knie et al.(2004)]{knie04} Knie K., Korschinek G., Faestermann T., Dorfi E.A., Rugel G., Wallner C., \prl, 93, 171103
\bibitem[Komiya et al.(2014)]{kom14} Komiya Y., et al., 2014, \apj, 783, 132
\bibitem[Koning et al.(2008)]{koning08} Koning  A.J., Hilaire S., Goriely S., 2008, Nucl. Phys., A810, 13
\bibitem[Koning et al.(2012)]{koning12} Koning A.J., Rochman D., Nuclear Data Sheets, 2012, 113, 2841
\bibitem[Kopecky \& Uhl(1990)]{kopecky90} Kopecky J, Uhl M., 1990, \prc, 41, 1941
\bibitem[Korobkin et al.(2012)]{korobkin12} Korobkin O., Rosswog S.,  Arcones A., Winteler C., 2012, \mnras, 426, 1940
\bibitem[Kyutoku \& Ioka(2016)]{Kyutoku16} K. Kyutoku, K. Ioka, 2016, eprint arXiv:1603.00467
\bibitem[Lattimer \& Schramm(1976)]{lattimer76} Lattimer J.M. \& Schramm D.N., 1976, \apj, 210, 549
\bibitem[Lattimer et al.(1977)]{lattimer77} Lattimer J.M., Mackie F., Ravenhall D.G., Schramm D.N., 1977, \apj, 213, 225
\bibitem[Lucy(1974)]{lucy74} Lucy L.B., 1974, AJ, 79, 745
\bibitem[Ludwig et al.(2016)]{ludwig16} Ludwig P., Bishop S., Egli R.,  Chernenko V., Deneva B., et al., 2016, Proc. Natl. Acad. Sci. U.S.A., accepted
\bibitem[Martin et al.(2015)]{martin15} Martin D., Perego A., Arcones A., Thielemann F.-K., Korobkin O., Rosswog S.,  2015, \apj, 813, 2
\bibitem[Matteucci et al.(2014)]{mat14} Matteucci  F., et al.,  2014, \mnras, 438, 2177 
\bibitem[Mennekens \& Vanbeveren(2014)]{mennekens14} Mennekens N. \& Vanbeveren D., 2014, \aap, 564, A134
\bibitem[Metzger et al.(2007)]{metzger07} Metzger B.D., Thompson T.A., Quataert E., 2007, \apj, 659, 561
\bibitem[Metzger et al.(2010)]{metzger10} Metzger B.D., Martinez-Pinedo G., Darbha S., et al., 2010, \mnras, 406, 2650
\bibitem[Mirizzi et al.(2016)]{mirizzi16} Mirizzi A., Tamborra I., Janka H.-T., Saviano N., Scholberg K., Bollig R., H\"udepohl L., Chakraborty S., 2016, La Rivista del Nuovo Cimento, 39, 1--2
\bibitem[M\"oller et al.(2003)]{moller03} M\"oller P., Pfeiffer B., Kratz K.-L., 2003, \prc, 67, 055802
\bibitem[M\"oller et al.(2016)]{moller16} M\"{o}ller P., Sierk A.J., Ichikawa T., Sagawa H., 2016, ADNDT, submitted
\bibitem[Myers \& Swiatecki(1999)]{ms99} Myers W.D., Swiatecki  W.J.,  1999, \prc, 60,  014606
\bibitem[Suzuki \& Nagataki(2005)]{suzuki05} Suzuki T.K. \& Nagataki S., \apj, 628, 914
\bibitem[Paul et al.(2001)]{paul01} Paul M., Valenta A., Ahmad I., et al., 2001, \apj, 558, L133
\bibitem[Perego et al.(2014)]{perego14} Perego A., Rosswog S., Cabez\'on R.M., Korobkin O., K\"appeli R., Arcones A., Liebend\"orfer M., 2014, \mnras, 443, 3134
\bibitem[Pruet et al.(2005)]{pruet05} Pruet J., Woosley S.E., Buras R., et al., 2005, \apj, 623, 325
\bibitem[Qian \& Woosley(1996)]{qian96} Qian Y.-Z. \& Woosley S.E., 1996, \apj, 471, 331
\bibitem[Radice et al.(2016)]{radice16} Radice D., Galeazzi F., Lippuner J., Roberts L.F., Ott C.D., Rezzolla L., 2016, eprint arXiv:1601.02426
\bibitem[Raisbeck et al.(2007)]{raisbeck07} Raisbeck G., Tran T., Lunney D., Gillard C., Goriely S., Waelbroeck C., Yiou F., 2007, Nuclear Instruments and Methods in Physics Research B, 259, 673
\bibitem[Roberts et al.(2010)]{roberts10} Roberts L.F., Woosley S.E., Hoffman R.D., 2010, \apj, 722, 954
\bibitem[Roberts et al.(2011)]{roberts11} Roberts L.F., Kasen D., Lee W.H., Ramirez-Ruiz E., 2011, \apjl, 736, L21
\bibitem[Roberts et al.(2016)]{roberts16} Roberts L.F., et al., eprint arXiv:1601.07942
\bibitem[Roederer et al.(2010)]{roederer10} Roederer I. U., Cowan J. J., Karakas A. I., Kratz K.-L., Lugaro M., Simmerer J., Farouqi K., Sneden C., 2010, ApJ, 724, 975
\bibitem[Roederer(2011)]{roederer11} Roederer I.U.,\ 2011, \apjl, 732, L17
\bibitem[Roederer et al.(2012)]{roederer12} Roederer I.U., et al.,  2012, ApJS, 203, 27
\bibitem[Rosswog et al.(1999)]{rosswog99} Rosswog S., Liebend\"orfer M., Thielemann F.-K., Davies M.B., Benz W., Piran T., 1999, \aap, 341, 499 
\bibitem[Schmidt \& Jurado(2012)]{gef12} Schmidt K.-H., Jurado  B., Phys. Procedia, 2012, 31, 147
\bibitem[Sekiguchi et al.(2015)]{seki15} Sekiguchi Y., Kiuchi K., Kyutoku K., Shibata M., 2015, \prd, 91, 064059
\bibitem[Shen et al.(2014)]{shen14} {Shen} S., {Cooke} R., {Ramirez-Ruiz} E., {Madau} P., {Mayer} L.,  {Guedes} J., 2014, arXiv:1407.3796
\bibitem[Sneden et al.(1996)]{sneden96} Sneden C., McWilliam A., Preston G.W., et al., 1996, ApJ, 467, 819
\bibitem[Sneden et al.(2003)]{sneden03} Sneden C., et al., 2003, ApJ, 591, 936
\bibitem[Sneden et al.(2008)]{sneden08} Sneden C., Cowan J. J., Gallino R., 2008, ARA\& A, 46, 241
\bibitem[Sneden et al.(2009)]{sneden09} Sneden C., Lawler J. E., Cowan J. J., Ivans I. I., Den Hartog E. A., 2009, \apjs, 182, 80
\bibitem[Sukhbold et al.(2015)]{sukhbold15} Sukhbold T., Ertl T., Woosley, S.E., Brown J.M., Janka H.-T., 2015, e-print arXiv:1510.04643
\bibitem[Tachibana et al.(1990)]{tachibana90} Tachibana T., Yamada M., Yoshida Y., 1990, Prog. Theor. Phys., 84,  641
\bibitem[Takahashi et al.(1994)]{taka94} Takahashi K., Witti J., Janka H.-Th., 1994, \aap, 286, 857
\bibitem[Takahashi \& Janka(1997)]{takahashi97} Takahashi K., Janka H.-Th., 1997,  in {\sl Origin of Matter and Evolution of Galaxies},  eds. T.~Kajino et al., (Singapore: World Scientific), p.~213
\bibitem[Thompson(2003)]{thompson03} Thompson T.A., 2003, \apjl, 585, L33
\bibitem[Thompson et al.(2001)]{thompson01} Thompson T.A., Burrows A., Meyer B.S. 2001, \apj, 562, 887
\bibitem[van de Voort  et al.(2014)]{voort14} {van de Voort} F., {Quataert} E., {Hopkins} P.~F., {Keres} D.,  {Faucher-Giguere} C.-A., 2015,  \mnras, 447, 140
\bibitem[Vangioni et al.(2015)]{vangioni15} Vangioni E., Goriely S., Daigne F., Fran\c cois P., Belczynski K.,  2015, \mnras, 455, 17 
\bibitem[Vlasov et al.(2014)]{vlasov14} Vlasov A.D., Metzger B.D., Thompson T.A., 2014, \mnras, 444, 3537
\bibitem[Wallner et al.(2004)]{wallner04} Wallner A., Faestermann T., Gerstmann U., Knie K., Korschine, G.,
Lierse C., Rugel G., 2004, New Astronomy Reviews 48, 145 
\bibitem[Wallner et al.(2015)]{wallner15} Wallner A., et al., 2015, Nature Communications, 6, 5956
\bibitem[Wanajo et al.(2011)]{wanajo11} Wanajo S., Janka H.-T., M\"uller B., 2011, \apjl, 726, L15
\bibitem[Wanajo et al.(2014)]{wanajo14} Wanajo S., et al., 2014, \apj, 789, L39
\bibitem[Wehmeyer et al.(2015)]{wehmeyer15} Wehmeyer B., Pignatari M., Thielemann F.-K., 2015, \mnras, 452, 1970
\bibitem[Westphal et al.(1998)]{westphal98} Westphal A.J., Price  P.B., Weaver B.A., \& Afanasiev V.G., 1998, Nature, 396, 50
\bibitem[Winteler et al.(2012)]{winteler12} Winteler C., K\"appeli R., Perego A., Arcones A., Vasset N., Nishimura N., Liebend\"orfer M., Thielemann F.-K., 2012, \apjl, L22
\bibitem[Witti et al.(1994)]{witti94} Witti J., Janka H.-Th., Takahashi K., 1994, \aap, 286, 841
\bibitem[Witze(2013)]{witze13} Witze A., 2013, Nature, 12797
\bibitem[Woosley et al.(1994)]{woosley94} Woosley S.E., Wilson J.R., Mathews G.J., Hoffman R.D., Meyer B.S.,
1994, \apj, 433, 229 
\bibitem[Xu et al.(2013)]{xu13} Xu Y, Goriely  S., Jorissen  A., Chen G.L., Arnould M., 2013, \aap, 549, A106

\end{thebibliography}
\end{document}